\documentstyle[twoside,fleqn,Espcrc2,epsf]{article}

\newcommand{\ggox}[3]{\gamma_{#1} \gamma_{#2} \otimes \xi_{#3}}
\newcommand{\gox }[2]{\gamma_{#1}             \otimes \xi_{#2}}

\title{Pion decay constant in quenched QCD with Kogut-Susskind quarks%
       \thanks{Presented by T. Kaneda}}

\author{JLQCD Collaboration:
        S.~Aoki%
        \address{Institute of Physics, University of Tsukuba,
        Tsukuba, Ibaraki 305-8571, Japan},
        M.~Fukugita%
        \address{Institute for Cosmic Ray Research, University of Tokyo,
        Tanashi, Tokyo 188-8502, Japan},
        S.~Hashimoto%
        \address{High Energy Accelerator Research Organization (KEK),
        Tsukuba, Ibaraki 305-0801, Japan},
        K-I.~Ishikawa$^{\rm c}$,
        N.~Ishizuka$^{\rm a,}$%
        \address{Center for Computational Physics, University of Tsukuba,
        Tsukuba, Ibaraki 305-8577, Japan},
        Y.~Iwasaki$^{\rm a,d}$,
        K.~Kanaya$^{\rm a,d}$,
        T.~Kaneda$^{\rm a}$,
        S.~Kaya$^{\rm c}$,
        Y.~Kuramashi$^{\rm c}$,
        M.~Okawa$^{\rm c}$,
        T.~Onogi%
        \address{Department of Physics, Hiroshima University,
        Higashi-Hiroshima, Hiroshima 739-8526, Japan},
        S.~Tominaga$^{\rm c}$,
        N.~Tsutsui$^{\rm e}$,
        A.~Ukawa$^{\rm a,d}$,
        N.~Yamada$^{\rm e}$,
        and
        T.~Yoshi\'e$^{\rm a,d}$}

\begin{document}

\begin{abstract}
We report results on the flavor breaking effect in the pion decay
constant $f_\pi$ calculated with the Kogut-Susskind quark action.
Numerical simulations are carried out at $\beta = 6.0$ and 6.2.  We find
that the use of non-perturbative renormalization factor leads to results
in the continuum limit well convergent among various KS flavors.  As the
best value for $f_\pi$ we obtain $f_\pi = 89(6)$ MeV from the conserved
axial vector current.
\end{abstract}

\maketitle

\section{Introduction}

The Kogut-Susskind (KS) quark action has the well-known feature that
SU(4) flavor symmetry is broken down to U(1) subgroup at finite lattice
spacing.  The restoration of full flavor symmetry toward the continuum
limit has been previously examined for pion mass\cite{Shar92,JLQC97}.
Here we extend the examination to the pion decay constant by comparing
results for various KS flavors to that in the U(1) channel for which the
renormalization constant equals unity.

This comparison is made both for perturbative\cite{pertur} and non-%
perturbative renormalization factors, the latter evaluated\cite{JLQC99}
with the method of Ref.~\cite{Mart95}.

Numerical simulations are carried out in quenched QCD at $\beta = 6.0$
and 6.2 employing $32^3\times 64$ and $48^3\times 64$ lattices.  Other
lattice parameters are summarized in Table~\ref{tab:parameter}.

\section{Formalism and simulation}

\begin{table}[tb]
\setlength{\tabcolsep}{0.2pc}
\newlength{\digitwidth} \settowidth{\digitwidth}{\rm 0}
\caption{Lattice parameters of simulation.}
\label{tab:parameter}
\begin{tabular}{cccccc}
\hline
$\beta$
    & $L^3\cdot T$
    & $m_q a$
    & \#conf.
    & $a^{-1}$(GeV)
    & $m_\pi / m_\rho$ \\
\hline
6.0 & $32^3\cdot 64$ & .010--.030 & 100 & 1.93(2) & .53--.72 \\
6.2 & $48^3\cdot 64$ & .008--.023 &  60 & 2.70(5) & .55--.75 \\
\hline
\end{tabular}
\end{table}

\begin{figure}[tb]
\centerline{\epsfxsize=65mm \epsfbox{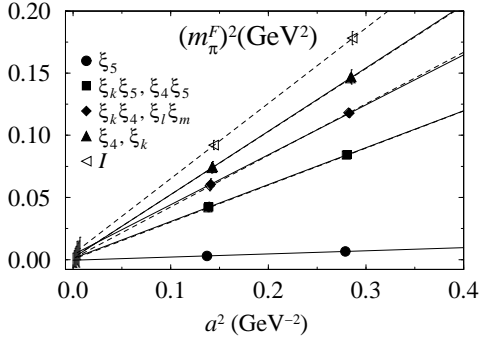}}
\caption{Continuum limit of pion masses.}
\label{fig:Pmass5}
\end{figure}
\begin{figure}[tb]
\centerline{\epsfxsize=65mm \epsfbox{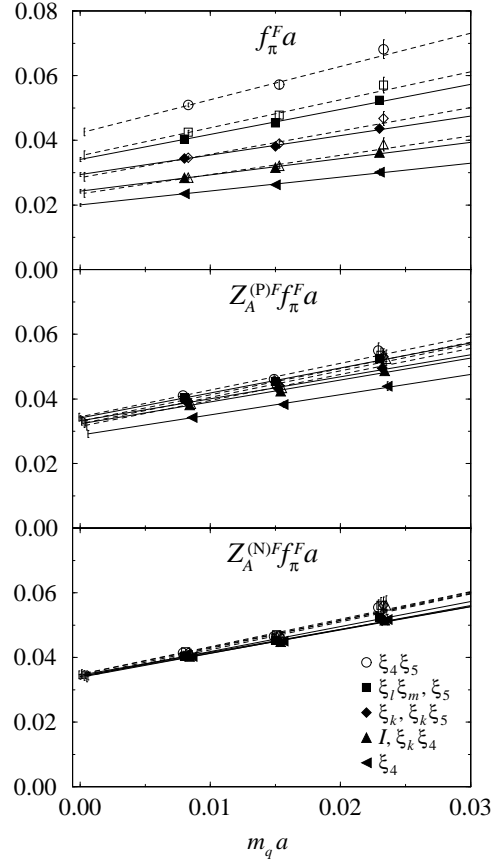}}
\caption{$f_\pi^F$ obtained with gauge-invariant axial vector current as
         functions of $m_q$ at $\beta = 6.2$ before renormalization
         (top), with perturbative (middle) and non-perturbative (bottom)
         renormalization factors.}
\label{fig:fpi1}
\end{figure}
\begin{figure}[htb]
\centerline{
\epsfxsize=65mm \epsfbox{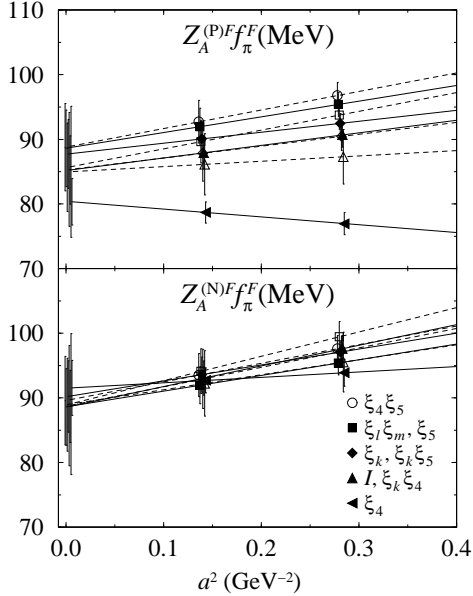}}
\caption{Continuum limit of $f_\pi^F$ with perturbative (top) and non-%
         perturbative (bottom) renormalization factor obtained with
         gauge-invariant axial vector current.}
\label{fig:fpi5-inv}
\end{figure}
\begin{figure}[htb]
\centerline{
\epsfxsize=65mm \epsfbox{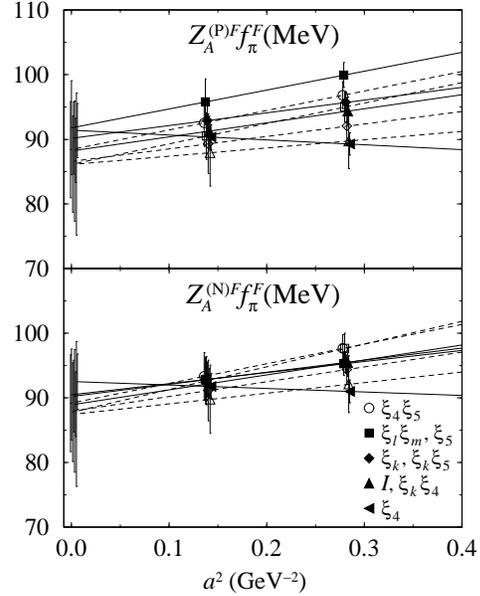}}
\caption{Same as Fig.~\ref{fig:fpi5-inv} obtained with gauge non-%
         invariant axial vector current.}
\label{fig:fpi5-non}
\end{figure}

In the hypercubic notation for the KS fermion fields, the axial vector
current in the KS flavor channel $F$ has the form
\begin{equation}
A_\mu^F =\overline{\phi} (\ggox{\mu}{5}{F})\phi.
\label{eq:axial}
\end{equation}
The fields $\overline{\phi}$ and $\phi$ in (\ref{eq:axial}) are
generally separated over a 4-dimensional hypercube.  We use both gauge-%
invariant and non-invariant operators, inserting the product of link
variables between $\overline{\phi}$ and $\phi$ for the former, and
employing the Landau gauge fixing to deal with the latter.

The one-loop perturbative results for the renormalization constants
$Z_A^F$ have been worked out in Ref.~\cite{pertur}.  We use tadpole-%
improved values employing the tadpole-improved $\overline{\rm MS}$
coupling with $q^* = 1 / a$; they will be denoted as $Z_A^{{\rm (P)}F}$.
For non-perturbative renormalization constants $Z_A^{{\rm (N)}F}$, we
take results of Ref.~\cite{JLQC99} at $(a p)^{2} = 1.0024$.

Quark propagators are calculated for 16 wall sources, each corresponding
to a corner of a hypercube.  These propagators are combined to form the
correlator $\langle A_\mu^F(t)\pi_W^F(0)\rangle$ for each flavor $F$,
where $\pi_W^F$ is the pion field for the wall source, thereby enhancing
signals\cite{Ishi94}.

We find the quark mass dependence of $(m_\pi^F)^2$, $f_\pi^F$ and vector
meson masses to be well described by a linear function.  Hence a linear
fit is employed for the chiral extrapolation in the quark mass $m_q$.
The scale is set by the $\rho$ meson mass in the spin-flavor channel
$\gox{k}{k}$.  Errors are estimated by the single elimination jackknife
procedure.

\section{Pion masses}

We show in Fig.~\ref{fig:Pmass5} values of $(m_\pi^F)^2$ at $m_q = 0$ as
a function of $a^2$.  The 16 KS flavors are grouped into 8 irreducible
representations (irreps) given by\cite{Golt86}
$\xi_5, \xi_4\xi_5, \xi_4, I$ (1-d irreps) and
$\xi_k\xi_5, \xi_k\xi_4, \xi_\ell\xi_m, \xi_k$ (3-d irreps).  We
observe that the 8 irreps form a degeneracy pattern
$\xi_5, (\xi_k\xi_5, \xi_4\xi_5), (\xi_k\xi_4, \xi_\ell\xi_m),
 (\xi_4, \xi_k), I$ at finite lattice spacing.  This feature was
initially observed numerically in Ref.~\cite{Ishi94}, and recently
theoretically explained in Ref.~\cite{LeeS99}.  We also see that non-%
zero values of $(m_\pi^F)^2$ for the channels other than $\xi_5$ vanish
quadratically in $a$ toward the continuum limit as expected.

\section{Pion decay constants}

In Fig.~\ref{fig:fpi1} we illustrate how the bare values of $f_\pi^F$
change under renormalization, taking results for gauge invariant current
at $\beta = 6.2$.  The bare values for the 8 irreps again form a
degeneracy pattern.  The pattern reflects the distance of
$\overline{\phi}$ and $\phi$ in (\ref{eq:axial}), and is different from
that of $m_\pi^F$.

The perturbative renormalization constants (middle frame in Fig.~%
\ref{fig:fpi1}) help to reduce the discrepancy among the bare values of
$f_\pi^F$.  It is clear, however, that the discrepancy is much more
reduced with the non-perturbative renormalization constants
(bottom frame).

In Fig.~\ref{fig:fpi5-inv} we plot the continuum extrapolation of the
pion decay constants obtained with gauge-invariant currents.  The top
figure shows results with perturbative renormalization factors, and the
bottom figure with non-perturbative factors.  Lines are quadratic fits
in $a^2$ following the $O(a^2)$ scaling violation expected for the KS
quark action.  Similar figures for the decay constants obtained with the
gauge non-invariant currents are shown in Fig.~\ref{fig:fpi5-non}.

The gauge-invariant current for the flavor $\xi_5$ does not require
renormalization, from which we obtain $f_\pi = 89(6)$ MeV in the
continuum limit.  This result is consistent with the experimental value
of 92.4(3) MeV\cite{PDG_98}; possible deviations due to quenching is not
visible within our 7\% error.

We observe in Figs.~\ref{fig:fpi5-inv} and \ref{fig:fpi5-non} that the
perturbative renormalization factors at one-loop order are not
sufficient to ensure restoration of flavor symmetry in the continuum
limit.  In contrast, the decay constants evaluated with the non-%
perturbative renormalization factors agree much better already at finite
lattice spacings, and their continuum limits are well convergent, the
central values coinciding with each other well within the errors.

\vspace*{10pt}
This work is supported by the Supercomputer Project No.45 (FY1999) of
High Energy Accelerator Research Organization (KEK), and also in part by
the Grants-in-Aid of the Ministry of Education (Nos. 09304029, 10640246,
10640248, 10740107, 10740125, 11640294, 11740162).  K-I.I is supported
by the JSPS Research Fellowship.


\begin{thebibliography}{9}
\bibitem{Shar92} S.R.~Sharpe {\it et al.},
                 Nucl. Phys. B (Proc. Suppl.) 26 (1992) 197.
\bibitem{JLQC97} JLQCD Collaboration: T.~Yoshi\'e {\it et al.},
                 Nucl. Phys. B (Proc. Suppl.) 53 (1997) 209.
\bibitem{pertur} D.~Daniel and S.N.~Sheard,
                 Nucl. Phys. B302 (1988) 471;
                 A.~Patel and S.R.~Sharpe,
                 Nucl. Phys. B395 (1993) 701;
                 N.~Ishizuka and Y.~Shizawa,
                 Phys. Rev. D49 (1994) 3519.
\bibitem{JLQC99} JLQCD Collaboration: S.~Aoki {\it et al.},
                 Phys. Rev. Lett. 82 (1999) 4392;
                 in preparation.
\bibitem{Mart95} G.~Martinelli {\it et al.},
                 Nucl. Phys. B445 (1995) 81.
\bibitem{Ishi94} N.~Ishizuka {\it et al.},
                 Nucl. Phys. B411 (1994) 875.
\bibitem{Golt86} M.F.L.~Golterman,
                 Nucl. Phys. B273 (1986) 663.
\bibitem{LeeS99} W.~Lee and S.R.~Sharpe, hep-lat/9905023.
\bibitem{PDG_98} Particle Data Group: C.~Caso {\it et al.},
                 Eur. Phys. J. C3 (1998) 1.
\end{thebibliography}
\end{document}